\documentclass[12pt]{article}

\usepackage{epsfig}
\usepackage{amssymb,amsmath}
\usepackage{slashed}

\newcommand{\bea}{\begin{eqnarray}}
\newcommand{\eea}{\end{eqnarray}}

 \makeatletter
\def\alt{\mathrel{\mathpalette\gl@align<}}
\def\agt{\mathrel{\mathpalette\gl@align>}}
\def\gl@align#1#2{\lower.6ex\vbox{\baselineskip\z@skip\lineskip\z@
\ialign{$\m@th#1\hfil##\hfil$\crcr#2\crcr\sim\crcr}}} \makeatother

\begin{document}
\begin{flushright}
BA-07-32
\end{flushright}
\vspace*{1.0cm}

\begin{center}
\baselineskip 20pt {\Large\bf Unparticle Physics And Gauge Coupling
Unification} \vspace{1cm}

{\large Ilia Gogoladze$^{a,}$\footnote{ On  leave of absence from:
Andronikashvili Institute of Physics, GAS, 380077 Tbilisi, Georgia.
\\ \hspace*{0.5cm} },
Nobuchika Okada$^{b,c}$ and Qaisar Shafi$^{a}$} \vspace{.5cm}

{\baselineskip 20pt \it
$^a$Bartol Research Institute, Department of Physics and Astronomy, \\
University of Delaware, Newark, DE 19716, USA \\
\vspace{2mm} $^b$Department of Physics, University of Maryland,
College Park,  MD 20742, USA \\
\vspace{2mm} $^c$Theory Division, KEK, Tsukuba 305-0801, Japan }
\vspace{.5cm}

\end{center}

\begin{abstract}
Unparticle physics from a hidden conformal sector can alter
 the evolution of the Standard Model (SM) gauge couplings
 via TeV scale threshold corrections.
We discuss how this may lead to gauge coupling
 unification at $M_{GUT}\approx 2 \times 10^{15}$ GeV -- $5 \times 10^{17}$ GeV
  without introducing new particles in the SM sector.
\end{abstract}
\thispagestyle{empty}

\newpage

\addtocounter{page}{-1}

\baselineskip 18pt


It has been recognized for a long time that with a canonical
normalization of 5/3 for $U(1)_Y$, the three SM gauge couplings fail
to unify at  without introducing new physics, which often means the
introduction of some new particles between the electroweak scale and
the unification scale $M_{\rm GUT}$. The minimal supersymmetric
standard model (MSSM) provides the most compelling example of this
approach with $M_{\rm GUT}\approx 2\times 10^{16}$ GeV
\cite{gauge7}. An alternative scenario in which no new particles are
introduced, can be realized from higher dimensional GUTs in which
the canonical normalization for $U(1)_Y$ gauge coupling is replaced
by a different choice, say 4/3, which leads to gauge coupling
unification at $M_{\rm GUT}$ close to $4 \times 10^ {16}$ GeV
\cite{unification}. In some other modifications of the SM involving
low energy supersymmetry plus additional 'matter' fields,
unification of the gauge couplings at the string scale ($\sim 5
\times 10^ {17}$ GeV) can be realized \cite{Dienes}.

In this paper we propose to achieve unification of the SM gauge
couplings by exploiting the so-called "unparticle physics" sector
introduced in \cite{Georgi:2007ek}. We find that threshold
corrections via unparticle physics at the TeV scale can succeed in
unifying the three SM gauge couplings with the conventional 5/3
normalization for $U(1)_Y$, and without introducing any new
particles. Depending on  the size of these threshold corrections for
each SM gauge coupling,
 the unification scale can vary by a few order of magnitudes
 around $10^{15}$ GeV.
These  corrections  may be revealed
 by precision  measurements of the SM gauge couplings
 around the TeV scale at future high energy collider experiments
 such as the LHC and the International Linear Collider (ILC).

The basic structure of the unparticle physics is as follows. First,
we introduce a coupling between a new SM singlet operator
 ($\cal{O}_{\rm UV}$) with dimension $d_{\rm UV}$
 and a SM operator  ${\cal O}_{\rm SM}$ with dimension $n$,
\bea
 {\cal L} = \frac{c_n}{M^{d_{\rm UV}+n-4}}
     \cal{O}_{\rm UV} {\cal O}_{\rm SM} ,
\eea
where $c_n$ is a dimensionless constant, and $M$ is the energy scale
 characterizing the new physics.
This new physics sector is assumed to become strong and
 conformal at some energy $\Lambda_{\cal U}$, and
 the operator $\cal{O}_{\rm UV}$ flows to the unparticle operator
 ${\cal U}$ with dimension $d_{\cal U}$.
In the low energy effective theory, we obtain an interaction given
by
 \bea {\cal L}=c_n
 \frac{\Lambda_{\cal U}^{d_{\rm UV} - d_{\cal U}}}{M^{d_{\rm UV}+n-4}}
 {\cal U} {\cal O}_{\rm SM}
\equiv
  \frac{1}{\Lambda^{d_{\cal U}+ n -4}}  {\cal U} {\cal O}_{\rm SM},
\eea
where the unparticle  dimension $d_{\cal U}$ is determined at the
scale  $\Lambda_{\cal U}$ (which is induced through
  dimensional transmutation), and $\Lambda$ is
 the (effective) cutoff scale of the low energy effective theory.
In this paper, we consider only a `` scalar'' unparticle and we
first focus on the interaction
 between the unparticle and the SM Higgs sector given by
 \cite{Fox:2007sy}, \cite{Chen:2007qr},
  \cite{Kikuchi:2007qd},
 \cite{Delgado:2007dx},
\bea
 {\cal L} = \frac{1}{\Lambda^{d_{\cal U}+ n - 4}}
 {\cal U} {\cal O}_{\rm SM}(H^\dagger H),
\eea
where $H$ is the SM Higgs doublet and
 ${\cal O}_{\rm SM}(H^\dagger H)$ is
a SM operator given as a function of the gauge invariant
 combination $H^\dagger H$.
Once the Higgs doublet develops a vacuum expectation value (VEV), a
tadpole term for the unparticle is induced, \bea {\cal
L}_{\slashed{\cal U}} =
 \Lambda_{\slashed{\cal U}}^{4-d_{{\cal U}}} {\cal U},
\label{tadpole}
\eea
where
 $ \Lambda_{\slashed{\cal U}}^{4-d_{{\cal U}}}=
  \langle {\cal O}_{\rm SM} \rangle/ \Lambda^{d_{\cal U}+n-4}$.
Through this tadpole term, the unparticle acquires a
  VEV and the conformal symmetry
 is broken \cite{Fox:2007sy}.
This VEV is given by
 \bea
  \langle {\cal U} \rangle
  \lesssim  \Lambda_{\slashed {\cal U}}^{d_{\cal U}} .
\eea With  the cutoff scale of the effective theory
  around a TeV, we expect
 $ \langle {\cal U} \rangle  \thickapprox 100$ GeV$-1$ TeV.

Next we introduce couplings  between the unparticle
 and the SM gauge bosons as follows \cite{Fox:2007sy}:
\bea
{\cal L}_{\cal U} =
 -\frac{\lambda_3}{4 g_3^2} \frac{\cal U} {\Lambda^{d_{\cal U}}}
      G^A_{\mu \nu} G^{A \mu \nu}
 -\frac{\lambda_2}{4 g_2^2} \frac{\cal U}{\Lambda^{d_{\cal U}}}
      F^A_{\mu \nu} F^{A \mu \nu}
 -\frac{\lambda_1}{4 g^{'2}}  \frac{\cal U} {\Lambda^{d_{\cal U}}}
      B_{\mu \nu} B^{ \mu \nu},
\label{Unp-gauge}
\eea
where $G^A_{\mu \nu}$, $F^A_{\mu \nu}$ and $B_{\mu \nu}$
 denote the field strengths for  the SM gauge group SU(3)$\times$SU(2)$\times$U(1)$_Y$,
 and $\lambda_i$ are  dimensionless coefficient of order unity or less.
For  $<{\cal U}> \neq 0$,  Eq.~(\ref{Unp-gauge}) leads
 to modifications of the gauge kinetic terms. (This is
 reminiscent  of some early work \cite{ref5} based on modification of gauge
 kinetic energy terms in GUTs). Thus
\bea {\cal L} = -\frac{1}{4 g_i^2} \left[
  1 + \lambda_i \frac{\langle {\cal U} \rangle}{\Lambda^{d_{{\cal U}}}}
 \right]  {\cal F}_{\mu \nu} {\cal F}^{\mu \nu}
\simeq -\frac{1}{4 g_i^2}
 \left(   1 + \epsilon_i    \right)
  {\cal F}_{\mu \nu} {\cal F}^{\mu \nu} ,
\label{modify} \eea where ${\cal F}_{\mu \nu}$ represents the
appropriate  SM  field strength, and \bea
 \epsilon_i \equiv
 \lambda_i \frac{\langle {\cal U} \rangle}{\Lambda^{d_{{\cal U}}}}
 \simeq \lambda_i
  \left(
  \frac{\Lambda_{\slashed{\cal U}}}{\Lambda }
 \right)^{d_{\cal U}}.
\eea
This modification can be interpreted as a threshold correction
 in the gauge coupling evolution across the scale
 $\langle {\cal U} \rangle^{1/d_{\cal U}} \sim \Lambda_\slashed{\cal U}$
 \cite{Fox:2007sy}.

For $\Lambda_\slashed{\cal U} \lesssim M_Z$,
 the $\epsilon$'s are  severely constrained by
 the current precision measurements
 on the fine structure constant \cite{Fox:2007sy}.
The evolution of the fine structure constant from
 low energy to the Z-pole ($M_Z$) is consistent with the SM,
 and the largest uncertainty arises from the fine structure constant
 measured at the $Z$ pole \cite{PDG},
\bea
 {\alpha}^{-1}_{em}(M_Z) &=&  127.918 \pm 0.019 .
\eea
This uncertainty (in the $\overline{\rm MS}$ scheme)
 can be converted to the constraint
\bea
\epsilon \lesssim 1.4 \times 10^{-4}.
\eea
Here, $\epsilon$  is an admixture of $\epsilon_1$
 and $\epsilon_2$ corresponding to the QED coupling.
In the following, we consider the case
 $\Lambda_\slashed{\cal U} > M_Z$, with the expectation that
 the gauge coupling evolution in this region
can be precisely measured in future experiments in order totest this
scenario.
 From Eq.~(\ref{modify}),  the threshold correction
 for each gauge coupling at $\Lambda_\slashed{\cal U}$ is given by
\bea
 \Delta g_i^2(\Lambda_\slashed{\cal U})
  = g_i^2(\Lambda_\slashed{\cal U}) \times \epsilon_i.
\label{threshold}
\eea
Since $\epsilon_i$ are   free parameters,
 we can regard the threshold corrections as theoretical ambiguities
 of the SM gauge couplings at $\Lambda_\slashed{\cal U}$
 associated with the unparticle physics. We are assuming here that
 the three $\epsilon_i$ are all distinct. The TeV scale unparticle
 physics is not ''aware" of the underlying grand unified theory at
  $M_{\rm GUT}$. One way to realize this is to consider a five
 dimensional GUT compactified on $S^1/Z_2$ such that only the SM gauge
 symmetry survives at one of the fixed points.
 The couplings in Eq.~(\ref{Unp-gauge})
 can be realized on this fixed point, with the conformal sector
 restricted to the 4D brane (fixed point).

Let us  now see how  the threshold corrections via unparticle
physics
 enable the SM gauge couplings to unify at $M_{\rm GUT}$
 without introducing any new particles or non-canonical
 normalization for the U(1)$_Y$ gauge coupling.
We employ two-loop renormalization group equations (RGE)
 for the running gauge couplings \cite{RGE},
\bea
 \frac{d g_i}{d \ln \mu} =  \frac{b_i}{16 \pi^2} g_i^3 +\frac{g_i^3}{(16\pi^2)^2}
\sum_{j=1}^3B_{ij}g_j^2, \label{gauge} \eea
 where $ \mu$ is the renormalization scale,
$g_i$ ($i=1,2,3$) are the SM  gauge couplings  and
\bea
 b_i = \left(\frac{41}{10},-\frac{19}{6},-7\right),~~~~~~~~
 { b_{ij}} =
 \left(
  \begin{array}{ccc}
  \frac{199}{50}& \frac{27}{10}&\frac{44}{5}\\
 \frac{9}{10} & \frac{35}{6}&12 \\
 \frac{11}{10}&\frac{9}{2}&-26
  \end{array}
 \right),
\eea
with
 $(\alpha_1, \alpha_2, \alpha_3)=(0.01681, 0.03354, 0.1176)$
 at the Z-pole ($M_Z$) \cite{PDG}.

Figure~1 shows the evolution of $g_i$ (more precisely of
$\alpha^{-1}_i=4 \pi/g^2_i$),
 after incorporating the threshold corrections,
with  $|\epsilon_i|  \leq  0.1$ as an example,
 at $\Lambda_\slashed{\cal U} =1$ TeV.
We show  in Figure~2 the result for $ \epsilon_1=-0.05$ and
$\epsilon_2 = \epsilon_3 = 0.1$. Unification of the three gauge
couplings  is achieved
 at $M_{\rm GUT}=2 \times 10^{15}$ GeV.
The parameter set in Figure~3,
 $\epsilon_1 = -0.18$,  $\epsilon_2=0.2$ and  $\epsilon_3 = 0.1$,
 realizes  gauge coupling unification
 at $M_{\rm GUT}=5 \times 10^{17}$ GeV (string scale).
Finally,  just as in the MSSM, gauge coupling unification
 at $M_{\rm GUT} =2 \times 10^{16}$ GeV
  is possible, as shown in Figure~4. Intermediate scale unification
  with unparticle has been discussed in \cite{yi}.

Baryon number is normally broken as a result of
 unification of quarks and leptons in GUT multiplets,
 and proton decay mediated by superheavy gauge bosons
 is a typical prediction. Non-observation of proton decay in current
experiments
 leads to the  bound $M_{\rm GUT} \gtrsim 2 \times 10^{15}$ GeV on
 $M_{\rm GUT}$
  \cite{Nath:2006ut}.

In summary, we have considered an unparticle sector
  which couples to the SM Higgs sector.
In this case, the Higgs VEV triggers not only
 the electroweak symmetry breaking but also
 the conformal symmetry breaking in the unparticle sector.
As a consequence, the unparticle naturally develops a VEV which,
through couplings   to the SM gauge bosons,
 works as threshold corrections in the evolution of
 the SM gauge couplings across the conformal symmetry breaking scale.
The threshold corrections alter the SM gauge couplings at high energies.
We have examined the possibility of gauge coupling unification
 through these threshold corrections with only the SM particle
 contents.
We have shown that this  unification can be successfully
 realized with  threshold corrections of order 10\%
 for the SM gauge couplings $\alpha_i$.
Such threshold correction  may
 be revealed through  precision measurements in future experiments
 at high energies.

\section*{Acknowledgments}
N.O. would like to thank the Particle Theory Group
 of the University of Delaware for hospitality during his visit.
This work is supported in part by
 the DOE Grant \# DE-FG02-91ER40626 (I.G. and Q.S.),
 and
 the Grant-in-Aid for Scientific Research from the Ministry
 of Education, Science and Culture of Japan,
 \#18740170 (N.O.).



%
\begin{figure}[t,width=12cm, height=8cm]
\begin{center}
{\includegraphics[height=7cm]{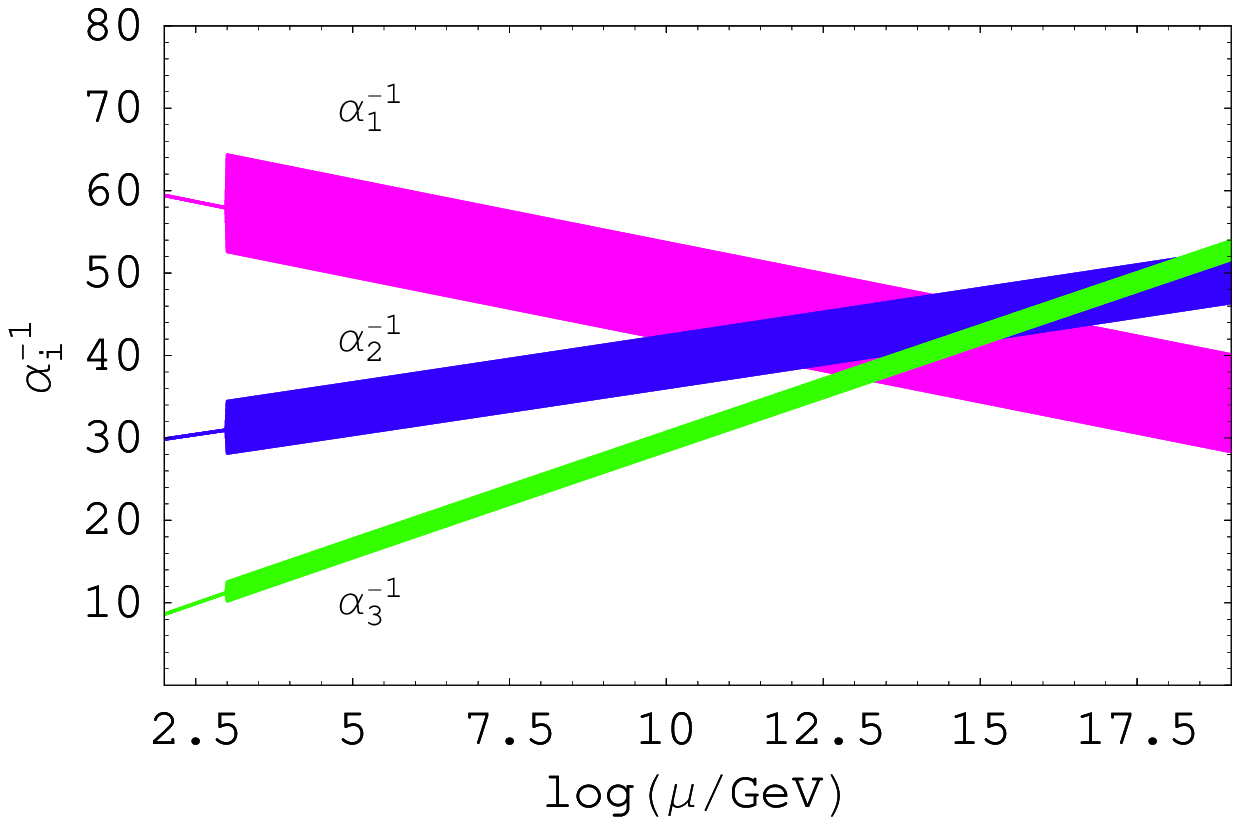} }
\end{center}
\caption{ \small Evolution of the three SM gauge couplings
$\alpha_i^{-1}$ with unparticle threshold corrections,
 $|\epsilon_i| \leq  0.1$,  at 1 TeV.
}
\end{figure}

\begin{figure}[t,width=12cm, height=8cm]
\begin{center}
{\includegraphics[height=7cm]{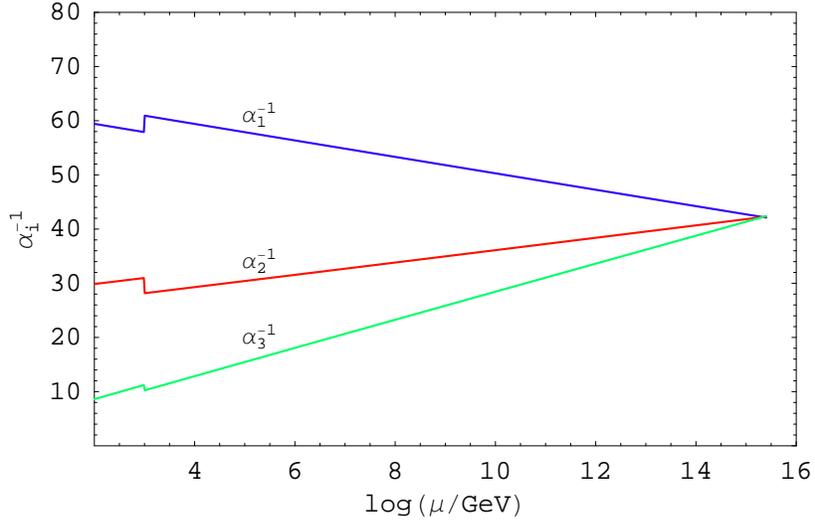} }
\end{center}
\caption{ \small Evolution of  $\alpha_i^{-1}$
 with unparticle threshold corrections at 1 TeV.
Here, $\epsilon_1 = -0.05$ and $\epsilon_2 = \epsilon_3 = 0.1$.
Unification occurs  at $ M_{\rm GUT}=2 \times 10^{15}$ GeV. }
\end{figure}

\begin{figure}[t,width=12cm, height=8cm]
\begin{center}
{\includegraphics[height=7cm]{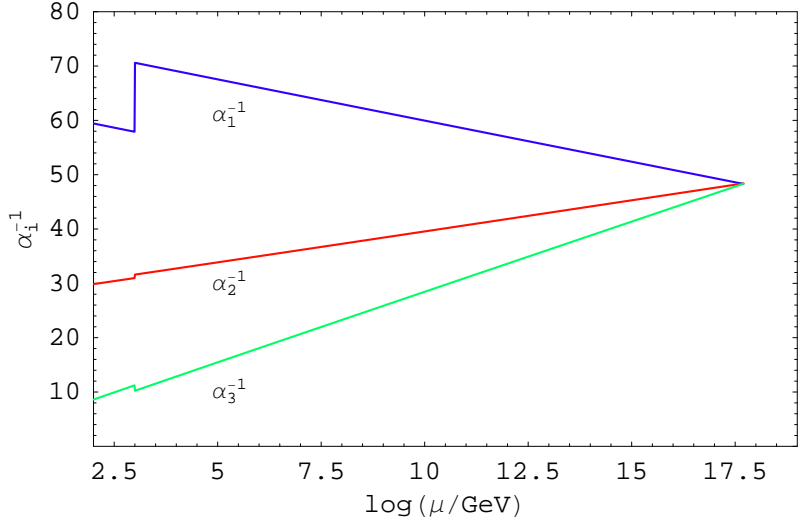} }
\end{center}
\caption{ \small Evolution of  $\alpha_i^{-1}$
 with unparticle threshold corrections at 1 TeV.
Here,
 $\epsilon_1 = -0.18$, $\epsilon_2 = -0.02$,  and $\epsilon_3 = 0.1$.
The three gauge couplings unify at
 $ M_{\rm GUT}=5 \times 10^{17}$ GeV.
}
\end{figure}

\begin{figure}[t,width=12cm, height=8cm]
\begin{center}
{\includegraphics[height=7cm]{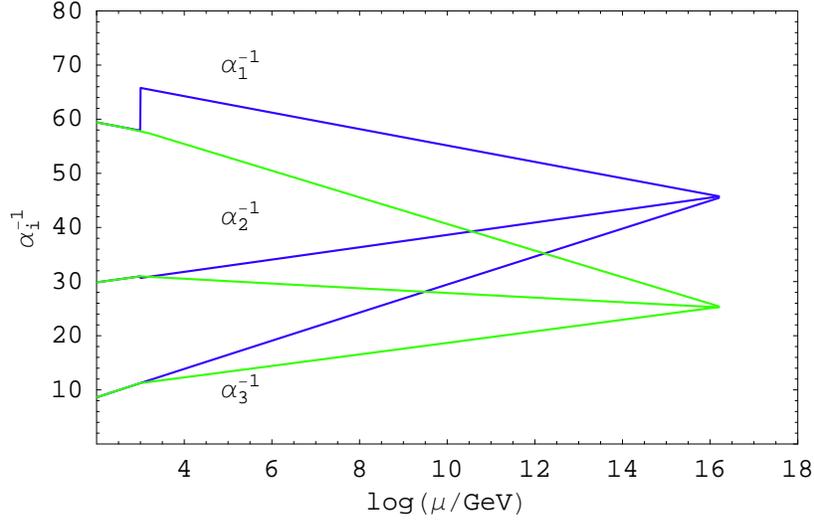} }
\end{center}
\caption{ \small Evolution of  $\alpha_i^{-1}$
 with unparticle threshold corrections,
 $\epsilon_1 = -0.12$, $\epsilon_2 = 0.01$ and $\epsilon_3 = 0$,
 at 1 TeV (blue lines).
Green lines show evolution of $\alpha_i^{-1}$ in
  MSSM  with the SUSY scale at 1 TeV.
}
\end{figure}
\end{document}